\documentclass[prb,aps,twocolumn,amsmath,amssymb]{revtex4}
\usepackage{graphicx}
\usepackage{dcolumn}
\usepackage{bm}

\begin{document}

\preprint{APS/123-QED}
\title{Electronic transport through a quantum dot network}

\author{August Dorn,$^1$ Thomas Ihn,$^1$ Klaus Ensslin,$^1$ Werner Wegscheider,$^2$ and Max Bichler$^3$}

\affiliation{
$^1$Solid State Physics Laboratory, ETH Z\"urich, 8093 Z\"urich, Switzerland\\
$^2$Institut f\"ur experimentelle und angewandte Physik, Universit\"at Regensburg, Germany\\
$^3$Walter Schottky Institut, Technische Universit\"at M\"unchen, Germany}

\date{\today}

\begin{abstract}
The conductance through a finite quantum dot network is studied as
a function of inter-dot coupling. As the coupling is reduced, the
system undergoes a transition from the antidot regime to the tight
binding limit, where Coulomb resonances with on average increasing
charging energies are observed. Percolation models are used to
describe the conduction in the open and closed regime and
contributions from different blockaded regions can be identified.
A strong negative average magnetoresistance in the Coulomb
blockade regime is in good quantitative agreement with theoretical
predictions for magnetotunneling between individual quantum dots.
\end{abstract}



\maketitle

\section{Introduction}

Arrays of insulating islands in two-dimensional electron systems
in the ballistic regime are often referred to as antidot lattices.
\cite{ensslin90,antidotrev} Features observed in the
magnetoresistance include commensurability peaks corresponding to
quasi pinned orbits around 1,4,9, ... antidos
\cite{weiss91,fleischmann92} and Aharonov-Bohm type oscillations
superimposed on the first commensurability peak.
\cite{weiss93,schuster94} The search for an artificial band
structure at B=0\,T is still ongoing.

Here we present measurements on a finite array with a very small
lattice constant of 120\,nm, where the global electron density can
be varied continuously with a metallic top gate. This allows us to
monitor the transition from an antidot to a quantum dot array,
that takes place when the electron density is reduced and the
constrictions between neighboring antidots enter the tunneling
regime. Our sample is special in the sense that the extremely
small lattice constant raises the charging energies to well
observable levels and ensures that the lattice enters the
tunneling regime well before the leads go insulating.

In the following we primarily focus on electronic transport in the
quantum dot network regime, which can be compared to conduction
through granular or disordered materials. Related systems include
arrays of metallic nano-crystals \cite{parthasarathy01,
parthasarathy04, black00, collier97, andres96}, layers of
semiconductor quantum dots \cite{ouyang03,morgan02}, porous
silicon \cite{hamilton98}, 3D arrays of semiconductor nanocrystals \cite{roest02,yu03}
and organic molecular crystals
\cite{schoonveld00}. However, unlike most experiments on
macroscopic samples, we are able to tune the inter dot coupling
continuously and resolve individual Coulomb resonances due to the
mesoscopic dimensions of our system. The local and global
properties of the network are investigated by measuring across
different terminals and phase coherence is probed by applying a
perpendicular magnetic field.

\section{Sample}

Starting with a high quality GaAs/AlGaAs heterostructure hosting a
two-dimensional electron system (2DES) 34\;nm below the surface,
we used AFM-lithography to define the nanostructure under study.
This patterning method relies on an atomic force microscope with a
conducting tip to locally oxidize the surface of a GaAs
heterostructure and thereby locally depleting the underlying 2DES
(see Refs \onlinecite{dorn02} and \onlinecite{fuhrer02} for
details). In this way a square lattice of $20 \times 20$
insulating islands with a lattice constant of a=120\;nm was
fabricated and enclosed by an insulating cavity with openings in
the corners that serve as current and voltage leads (see inset
Fig.1). Later a TiAu top gate was evaporated over the entire
structure using a shadow mask technique.

\begin{figure}[b]
\begin{center}
\includegraphics[width=3in]{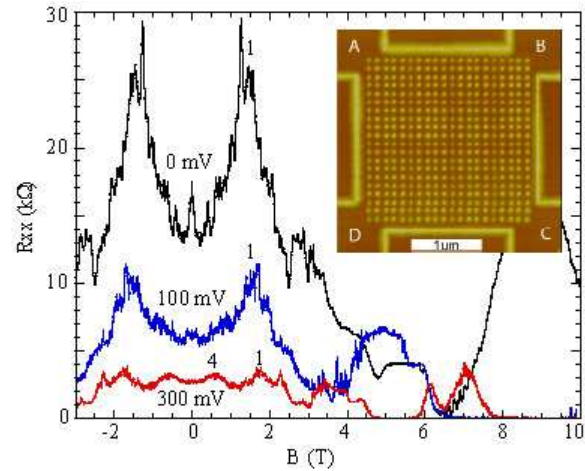}
\caption{
 Magnetoresistance measured from A to C across diagonal 1
at different top gate voltages at T=90\,mK, commensurability peaks
around 1 and 4 antidots are marked. Inset: AFM-micrograph of the
antidot lattice and the enclosing cavity. Bright regions are
oxidized and correspond to depletion in the underlying
two-dimensional electron
system.}\label{fig1a}\end{center}\end{figure}

\section{Measurements}

At high electron densities clear commensurability peaks around 1
and 4 antidots appear with superimposed ballistic conductance
fluctuations (see Fig. \ref{fig1a}), comparable to measurements
taken on a similar sample with a larger lattice constant by
Schuster et al. \cite{schuster94}. This indicates that despite the
small period, we still have a very symmetric two-dimensional
potential modulation. It is worth noting, that pronounced
commensurability maxima in the magnetoresistance even occur for
values exceeding the resistance quantum.

\begin{figure}[t]
\begin{center}
\includegraphics[width=3in]{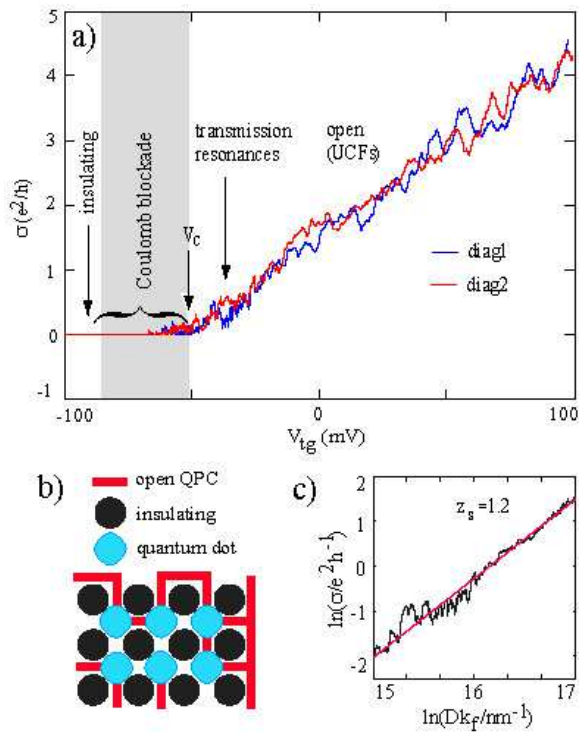}
\caption{ (a) Conductance as a function of top gate voltage across
diagonal 1(A-C) and diagonal 2(B-D) at T=90\,mK. Bright regions
are oxidized and correspond to depletion in the underlying 2DES.
 (b) Schematic of the lattice illustrating quantum dot formation and the bond percolation model.
 (c) Double logarithmic plot of the averaged conductance over
 $\Delta$$k_{f}=k_{f}-k_{fpc}$ with $k_{fpc} = k_{f}$ at $P_{c}$. The slope of the
 linear fit yields a value of $\zeta _{\sigma}=1.2$.
}\label{fig1b}\end{center}\end{figure}
\begin{figure}[t]
\begin{center}
\includegraphics[width=3in]{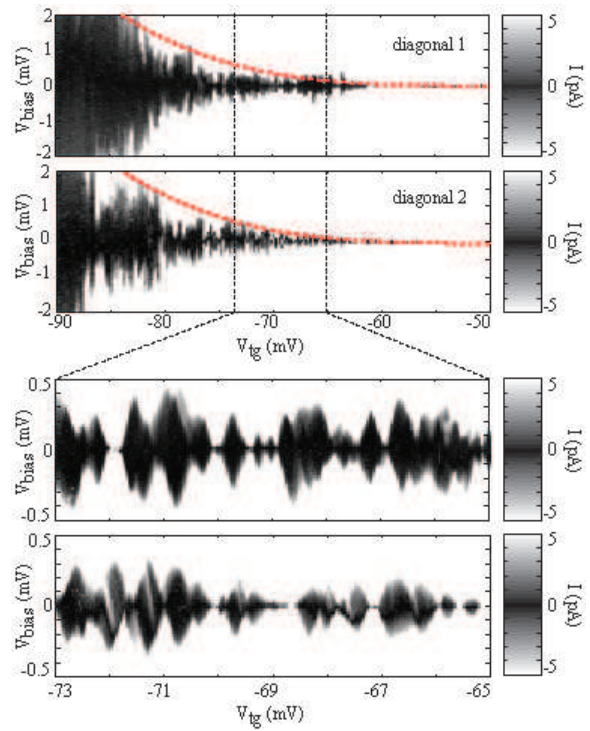}
\caption{ Current as a function of top gate and bias voltage.
White regions represent absolute current values above 5\;pA. The
upper two graphs are overview plots for diagonals 1 and 2, dashed
lines are guides to the eye and correspond to a scaling exponent
of about 3 with respect to $k_{F}$. The lower two graphs are
blowups showing well defined Coulomb diamonds. All measurements
were taken at a He bath temperature of 90\;mK at B\;=\;0\;T.
}\label{fig2}\end{center}\end{figure}

By applying suitable voltages to the top gate electrode, the
electron sheet density in the 2DES can be tuned from about 2 to
$5.5 \times 10^{15}$\;m$^{-2}$. The carrier density in the lattice
is about $1.5 \times 10^{15}$\;m$^{-2}$ lower than in the
unpatterned 2DES as determined from the Shubnikov--de Haas effect.
Since the resistance in the leads is smaller than $h/2e^{2}$ down
to top gate voltages below -100\;mV it can be neglected in all
measurements discussed here. Figure 2(a) shows the conductivity
measured across diagonal 1 from corner A to C and across diagonal
2 from corner B to D as a function of top gate voltage at
T=90\;mK. As the voltage is lowered, the electron sheet density
and the conductivity decrease until the Coulomb blockade regime is
reached. This transition takes place at a top gate voltage of
about $-50.8\;$mV, marked by $V_{c}$ in Fig. 2(a). Towards even
lower voltages a series of on average decreasing Coulomb peaks is
observed until conduction completely ceases below about $-90$\;mV.
It is worth pointing out, that the conductivity across both
diagonals is very similar over the entire range of top gate
voltages studied aside from mesoscopic fluctuations of the
conductance caused by interference and interaction. This indicates
the high symmetry and homogeneity of our sample. We would like to
stress, that the physics in the ballistic antidot regime at high
electron densities is in marked contrast to the quantum dot
network regime close to and below $V_{c}$. In particular the
linear transport characteristics of an antidot lattice and the
classical ballistic trajectories responsible for the
commensurability maxima give way to nonlinearities owing to
Coulomb charging and magnetotunneling between individual localized
states. During this transition the conductivity changes by several
orders of magnitude.



In order to gain more insight into the electronic properties of
the Coulomb blockade regime, we measured the current as a function
of top gate and bias voltage. As can be seen in Fig. 3, blockade
is lifted at sufficiently high bias voltages and clear `Coulomb
diamonds' are resolved. In contrast to analogous measurements on
single quantum dots, overlapping diamonds as well as streches in
top gate voltage without Coulomb blockade are observed. This
indicates the formation of a network with blockaded regions
connected in series and in parallel. The charging energy for the
individual resonances can be determined from the bias voltage
maxima $V_{max}$ at the tips of the Coulomb diamonds according to:

\begin{equation}
E_{charging}=eV_{max}= \frac{e^{2}}{C} + \Delta_{N} ,
\end{equation}

where $C$ is the capacitance and $\Delta_{N}$ is the quantum
mechanical single-particle energy spacing for the
N$^{\textrm{th}}$ state. Neglecting $\Delta_{N}$, which makes a
contribution of about 10\% to the total energy, and applying a
plate capacitor model, the area of the blockaded regions can be
determined using:

\begin{equation}
C = \epsilon_{0}\epsilon_{GaAs} \frac{A_{CB}}{d}
\end{equation}

where $A_{CB}$ is the area of the Coulomb blockaded region and $d$
is the distance between 2DES and top gate. This leads to an
average size of about 30 unit cells (or dots) at $V_{tg}=-60$\;mV
and about 4 unit cells at $V_{tg}=-80$\;mV, if the diameter of the
insulating discs is set to 80\;nm based on the oxide height
profile.

\section{Percolation Analysis}

In the following we apply percolation theory \cite{stauffer92} to
analyze these findings. This can be done by considering the
conductance of the entire lattice as being dominated by the
constrictions between neighboring insulating discs forming quantum
point contacts (QPCs). The area enclosed by four insulating
islands can then be viewed as a quantum dot or artificial atom
with four terminals connecting it to its nearest neighbor quantum
dots (see Fig. 2(b)). As the electron sheet density is reduced,
the conductivity of the QPCs decreases until the last channel
pinches off and the QPC goes insulating. Owing to small
inhomogeneities inherent to the fabrication process and the
presence of stray background charges, the closing of the QPCs will
be a statistical process. In a classical picture, the QPCs can
then be viewed as either conducting or broken resistors on a
square bond-percolation lattice.  Since the conductance staircase
of a QPC can be approximated to be linear with respect to the
Fermi wavenumber $k_{F} = \sqrt{2 \pi n(V_{tg})}$, where
$n(V_{tg})$ is the top gate voltage dependent electron sheet
density in the lattice, this parameter is a natural choice for
further analysis. In addition we identify the percolation
threshold $P_{c}$ for the open regime with $V_{c}=-50.8\;$mV and
assume the fraction of conducting bonds, or QPCs, $P$, to be
proportional to $\Delta k_{F}=k_{F}(V_{tg})-k_{F}(V_{c})$. The
exact relation depends on the width distribution of the QPCs, but
close to the percolation threshold $P_{c}=0.5$, \cite{stauffer92}
where half of the bonds are expected to be insulating, this is a
reasonable assumption. This suggests a percolation transition with
a characteristic scaling behavior of the type: \cite{stauffer92}

\begin{equation}
\sigma \propto (\Delta k_{F}) ^{\zeta}, \quad  \Delta
k_{F}=k_{F}(V_{tg})-k_{F}(V_{c}) ,
\end{equation}

where $\zeta$ is a critical exponent corresponding to a specific
quantity. From a double logarithmic plot of the conductance
averaged across both diagonals as a function of $\Delta k_{F}$, we
extract a conductivity scaling exponent $\zeta_{\sigma}=1.2 \pm
0.2$ in the open regime (Fig. 2(c)). This can be compared with the
calculated value of $1.32 \pm 0.02$ (Ref.
\onlinecite{bernasconi78}) for bond percolation in a classical
square random resistor network, if phase coherence is neglected.
Theoretical work describing a similar scenario has also been set
forth by Meir \cite{meir99} in a model for the metal insulator
transition in two dimensions.

\begin{figure}[b]
\begin{center}
\includegraphics[width=3in]{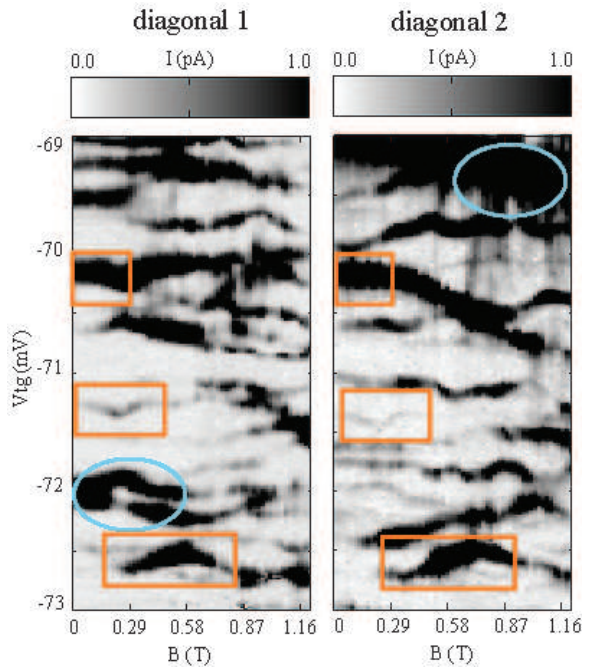}
\caption{ Conductance as a function of top gate voltage and
magnetic field across diagonals 1 and 2 at a He bath temperature
of 90\;mK. Boxes highlight similar features associated with the
same blockaded regions, while ovals mark features unique to one
diagonal.}\label{fig3}\end{center}\end{figure}

In the Coulomb blockade regime, we follow the `links, nodes, and
blobs model' introduced by Stanly \cite{stanly77} and Coniglio
\cite{coniglio82}. In this picture, the spanning network is
decomposed into multiply connected `blob bonds' and `dangling
bonds' forming `blobs', that are in turn linked by individual
`cutting' bonds. In our system the blobs correspond to clusters of
strongly coupled quantum dots that constitute the blockaded
regions, while the `cutting bonds' act as tunneling links. As the
electron density is reduced, more and more QPCs pinch off and the
clusters, or blobs, shrink, leading to higher charging energies.
The envelope functions in Fig. 3 correspond to a size scaling
exponent of about $\zeta_{CB} = 3 \pm 1$ as a function of $-\Delta
k_{F}$ \footnote{In the Coulomb blockade regime, $k_{F}$ was
determined by extrapolating the top gate voltage / electron sheet
density dependence from the open regime.} which is comparable to
theoretical calculations for the blob size scaling exponent
$\xi_{B}=2.06-2.16$ (Ref.\onlinecite{coniglio82}) and the mean
cluster size $\gamma=43/18\approx2.4$. \cite{stauffer92}

In order to make this interpretation more consistent, it might be
more appropriate to consider a two stage process. First a QPC goes
from open $(\sigma>2e^{2}/h)$ to tunneling $(\sigma < 2e^{2}/h$),
before becoming practically insulating $(\sigma \sim 0)$. The
first stage describes the transition from the open to the Coulomb
blockaded regime, while the second one induces the transition from
the Coulomb blockaded to an insulating state. The spanning cluster
above $k_{F}(V_{c})$ then consists of open QPCs, while the cutting
bonds in Coulomb blockade are formed by QPCs in the tunneling
regime. In either case the interpretation of the critical
exponents should be viewed as tentative owing to the experimental
uncertainty concerning the fraction $P$ of open (tunneling) QPCs.
We also point out, that a precise theoretical understanding of
Coulomb blockade scaling that includes effects from clusters
connected in series and in parallel is still outstanding.


Individual clusters of quantum dots in the network can be
monitored by measuring the shift of the Coulomb peaks as a
function of magnetic field across both diagonals (Fig. 4). Since
the magnetic field dependent energy variation of a quantum state
is related to the exact shape and symmetry of its wave function,
this variation can be regarded as a fingerprint of a specific
cluster. Similar features in Fig. 4 can be attributed to the same
cluster being traversed by current flow across both diagonals,
whereas differing features originate from clusters predominantly
probed by transport across one of the two diagonals. This
demonstrates, that transport is not dominated by a single quantum
dot or small cluster close to one of the leads.

\section{Phase Coherence}

Phase coherence across the spanning network was investigated by
measuring the magnetoconductance as a function of top gate voltage
(Fig. 5). The averaged magnetoconductance in the blockaded regime
shows an increase with magnetic field by a factor of approximately
3, which is significantly higher than the value of 4/3 predicted
by Alhassid \cite{alhassid00} and measured by Folk et al.
\cite{folk01} for single quantum dots. This increase occurs on a
magnetic field scale that corresponds to one flux quantum per unit
cell and is consistent with a magnetotunnling effect proposed by
Raikh and Glazmann \cite{raikh95} between elliptical `electron
lakes'. \footnote{This result should also hold for two chaotic
quantum dots.} They predict a low field magnetoresistance of the
form:

\begin{equation}
\frac{\delta R(B)}{R(0} \approx -\frac{B^{2}}{B_{0}^{2}}
\label{tunneling}
\end{equation}

where $B_{0}$ depends on the details of the tunnel barrier, but is
typically of the order of  $h/e \cdot 1/d_{1}d_{2}$, $d_{1}$ and
$d_{2}$ being the semiaxes of the two electron lakes. A best fit
to the magnetoconductance averaged with respect to top gate
voltage in the Coulomb blockade regime (dashed curve in Fig. 5(b))
yields a value of 2.2\;T for $B_{0}$. This corresponds to an
average radius of about 40\;nm, which is compatible with a single
quantum dot confined to a unit cell. Evidence for this effect has
also been reported by Voiskovskii and Pudalov
\cite{voiskovskii95}, however without resolving the increase in
conductivity of individual Coulomb peaks. Since Ref.
\onlinecite{raikh95} assumes perfect coherence within an electron
lake and from the characteristic field scale of a flux quantum
through a unit cell, we conclude, that the average dot radius can
be considered as a lower bound for the phase coherence length in
the Coulomb blockade regime. Remnants of this effect are still
visible in the open regime (Fig. 5(d)), but weak localization is
more prominent. From the dip around $\pm 5\;$mT we extract a phase
coherence length of about 300\;nm in the open regime.

\begin{figure}[t]
\begin{center}
\includegraphics[width=3in]{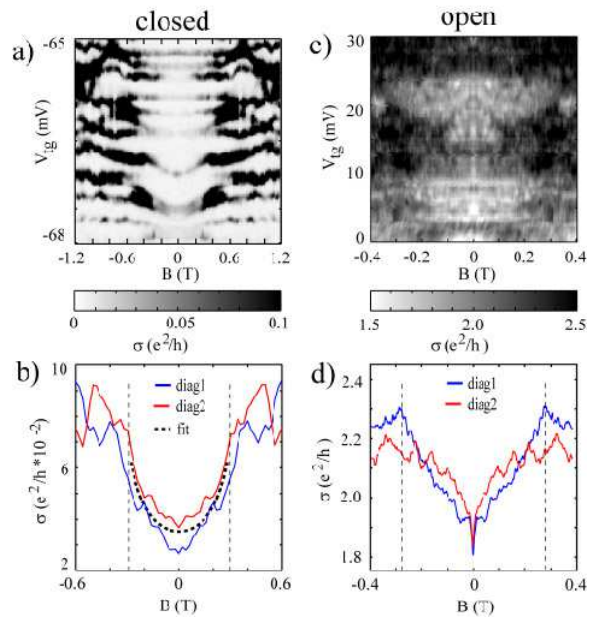}
\caption{ (a) Conductance as a function of top gate voltage and
magnetic field in the Coulomb blockade regime at
$\textrm{T}=90$\,mK. A clear positive magnetoconductance for
individual Coulomb peaks is observed. (b) Averaged
magnetoconductance between -54\;mV and -70\;mV in steps of
0.02\;mV. Dashed lines mark a flux quantum through the unit cell
(290\;mT) and a fit using equation \ref{tunneling}. (c)
Conductance as a function of top gate voltage and magnetic field
in the open regime. (d) Averaged magnetoconductance between 30 and
0\;mV in steps of 0.25\;mV. Dashed lines mark a flux quantum
through the unit cell (290\;mT).
}\label{fig4}\end{center}\end{figure}


In conclusion we have presented measurements on a multiply
connected multi-terminal quantum dot network with tunable
inter-dot coupling. Our sample is at a mesoscopic scale, where
collaboratively Coulomb blockaded regions can be discriminated and
related to macroscopic properties. For strong coupling close to
the percolation threshold, a classical random resistor network
model with superimposed quantum fluctuations can be applied until
charge quantization becomes important in the tunneling regime. For
weak coupling, Coulomb blockade dominates resulting in the
theoretically predicted \cite{middelton93, kaplan03, muller00} and
experimentally observed \cite{duruoz95, kurdak98} insulating state
for T$\rightarrow$0 with current onset above a bias voltage
threshold and hopping transport at elevated temperatures. A strong
parabolic decrease in average magnetoresistance in the Coulomb
blockade regime around B=0\;T, is in good quantitative agreement
with theoretical predictions by Raikh and Glazman \cite{raikh95}.
These findings complement reports by Wiebe et al. \cite{wiebe03},
who performed scanning tunneling experiments, Ilani et al.
\cite{ilani01}, who obtained local potential information, and by
Eytan et al. \cite{eytan98} who did scanning near field optical
microscopy on two-dimensional percolating systems. We are able to
quantitatively distinguish two length scales that are of
fundamental importance for transport in the weak coupling regime.
The first represents the decreasing area of the Coulomb blockaded
clusters while the second one describes wavefunction localization
to within the individual unit cells. Our measurements can also
serve as an intuitive picture for the formation of the so-called
`Coulomb gap', \cite{shklovskii84} that opens up around the Fermi
energy as a function of electron localization and Coulomb
interactions.

We thank Y. Gefen, R. Leturcq, C. Marcus, and D. Pfannkuche for
valuable discussions. This work was supported by the
Schweizerischer Nationalfonds.

\end{document}